\documentclass[12pt]{article}

\def\ai{{a^{}_i}}
\def\aix{{a^+_i}}
\def\aim{{a^-_i}}
\def\ajx{{a^+_j}}
\def\aj{{a^{}_j}}
\def\ajm{{a^-_j}}
\def\akx{{a^+_k}}

\def\akm{{a^-_k}}
\def\NN{{\bf N}}
\def\be{\begin{equation}}
\def\ee{\end{equation}}
\newtheorem{definition}{Definition}
\newtheorem{theorem}{Theorem}
\begin{document}

\title{NUMBER OPERATOR ALGEBRAS AND GENERALIZATIONS OF SUPERSYMMETRY}

\author{Fabien Besnard
\footnote{(besnard@math.jussieu.fr)}
}




\maketitle
\begin{abstract}
Several quantum systems with exotic statistics have been used in the last few years to extend supersymmetry. We show how all these systems fit into the picture of what we call ``Number Operator Algebras''.
\end{abstract}

\section{Introduction}
The algebra of $N=1$ supersymmetry is generated by $Q$ (the supercharge) and its conjugate $Q^+$, subject to the relations :

\be
Q^2={Q^+}^2=0\label{susy1}
\ee
\be
\{Q,Q^+\}=H\label{susy2}
\ee
where $\{Q,Q^+\}=QQ^++Q^+Q$ is the anti-commutator. These relations immediately imply :
\be
[H,Q]=[H,Q^+]=0\label{susy3}
\ee
where $[H,Q]=HQ-QH$ is the commutator. This algebra can be obtained in the following way : let $f^{\pm}$ (respectivley $b^{\pm}$) be fermionic (respectively bosonic) operators, generating the algebra $F$ (respectively $B$). Then let $Q:=b^+\otimes f^-$ and $Q^+=b^-\otimes f^+\in B\otimes F$. Note that $Q$ acts by exchanging a fermion state for a boson state on the Hilbert space tensor product of the respective Fock representation of one boson and one fermion, so that equations (\ref{susy3}) mean that the quantum equations of evolution of the system are invariant under such an exchange. We embed $F$ (resp. $B$) into $B\otimes F$ by $f^\pm\mapsto 1\otimes f^\pm$ (resp. $b^\pm\mapsto b^\pm\otimes 1$) so that in the following we allow ourselves to write $b^\pm$ and $f^\pm$ for the corresponding elements of $B\otimes F$.

By definition of the tensor product algebra we have :

\be
[b^\pm,f^\pm]=0
\ee

And by definition of $B$ and $F$ we have :
\be
[b^-,b^+]=1\label{boson}
\ee
and
\be
{f^-}^2={f^+}^2=0\label{fermion1}
\ee
\be
\{f^-,f^+\}=1\label{fermion2}
\ee

It is immediate to verify that equations (\ref{susy1}), and (\ref{susy2}) are satisfied. Note that $H=b^+b^-+f^+f^-:=H_b+H_f$.

We will say that a generalization of $N=1$ SUSY is a family of algebras, depending on one or several parameters so that (\ref{susy1}) and (\ref{susy2}) are recovered for a certain value of the parameters. Moreover, since (\ref{susy3}) expresses the invariance of the Hamiltonian under symmetry, we ask for such a commutation relation to be true between the generalized supercharge(s) and Hamiltonian. 
To obtain such a generalization a natural recipe is to start from a generalization of equations (\ref{boson})-(\ref{fermion2}) and define the supercharge accordingly. That is to say : generalized SUSY intertwines two particles with different statistics, at least one of which will have exotic statistics. We begin this paper by reviewing the constructions of this kind that have been made so far : first orthoSUSY, then paraSUSY. Then we will look at the link with Number Operator Algebras. These are algebras generated by creation and annihilation operators, in which number operators exist. We will give a classification theorem, under some restrictive assumptions. This theorem will allow us to rediscover many of the algebras seen before in the first two parts. In the last part we use a particular number operator algebra to define a new generalization of supersymmetry.

\section{Orthosupersymmetry}

Orthofermions have been first defined by Mishra and Rajasekaran\cite{mishraraja1,mishraraja2,mishraraja3,kharemishra1}. An orthofermion is a particle whose creation and annihilation operators $\ai^\pm$ satisfy :

\be
\aim\ajm=\aix\ajx=0\label{ortho1}
\ee

\be
\aim\ajx+\delta_{ij}\sum_k\akx\akm=\delta_{ij}\label{ortho2}
\ee

The index $i$ stands for an internal degree of freedom that could be called ``orthospin'' and runs from one to $n$. The algebra generated by the $\ai^\pm$ satisfying the previous relations will be subsequently denoted $E_n$ and called orthofermionic algebra. For $n=1$ we recover the usual fermionic algebra. Note that the $N_k=\akx\akm$ are number operators and $H_o=\sum_k N_k$ is the Hamiltonian.  The relations (\ref{ortho1}) and (\ref{ortho2}) entail that the orthofermion algebra is linearly spanned by $1$, $a^-_1$,\ldots, $a^+_n$, and the $\aix\ajm$ for $1\leq i,j\leq n$. That $E_n$ is indeed freely spanned by those vectors can be seen by noticing that the relations are {\it confluent} under the deglex-ordering $a^+_1<\ldots<a^+_n<a^-_1<\ldots<a^-_n$ and by using Bergman's diamond lemma. In short, this means that replacing every occurence of $\aim\ajm$ and $\aix\ajx$ by $0$ and $\aim\ajx$ by $\delta_{ij}(1-\sum_k\akx\akm)$ in any monomial in the generators is a well defined (non-ambiguous) procedure that always terminates (see Ref. \cite{berg} or \cite{bes1} for details).
Thus this algebra is $(n+1)^2$-dimensional. The Fock representation is easily defined and is faithful\cite{mosta1,bes4}, so that $E_n$ is isomorphic to the algebra of square matrices of dimension $n+1$.

Following our general recipe given in the introduction, we can look now for the algebra satisfied by $Q_i:=b^+\aim$ and $Q_i^+:=b^-\aix$ where $b^\pm$ are ordinary boson operators. The following equations are easily verified\cite{kharemishra1} :

\be
Q_iQ_j=Q_i^+Q_j^+=0\label{orthosusy1}
\ee

\be
Q_iQ_j^++\delta_{ij}\sum_kQ_k^+Q_k=\delta_{ij}H\label{orthosusy2}
\ee

with $H=H_b+H_o$ where $H_b=b^+b^-$ is the bosonic Hamiltonian\footnote{up to an additive constant that is irrelevant for our purpose}. Thus :
$$[Q_i,H]=[b^+\aim,H_b]+[b^+\aim,H_o]=-b^+\aim+b^+\aim=0$$ 
And by conjugation we get $[Q_i^+,H]=0$. Note that we could also have used equations (\ref{orthosusy1}) and (\ref{orthosusy2}) only without writing the explicit form of $H$\cite{tomiya}. Thus a quantum system is said to have orthosupersymmetry of order $n$ if there exists supercharges $Q_i$ so that (\ref{orthosusy1}) and (\ref{orthosusy2}) are satisfied. Then it is automatic that $[H,Q_i]=[H,Q_i^+]=0$.

\section{Parasupersymmetry}

The paraSUSY algebra of order $2j$, with $j\in{1\over 2}\NN$, is given by\cite{tomiya} :
\be
Q_j^{2j}Q_j^++Q_j^{2j-1}Q_j^+Q_j+\ldots+Q_j^+Q_j^{2j}=\alpha_jQ_j^{2j-1}H\label{parasusy1}
\ee
with $\alpha_j={2\over 3}j(j+1)(2j+1)$,
\be
Q_j^{2j+1}=0\label{parasusy2}
\ee
\be
[Q_j,H]=0
\ee
plus the adjoint equations. It can be realized, using our general recipe, by taking $Q_j:=b^+f_j$, where $b^+$ is a boson creator and $f_j$ a parafermion annihilator of order $2j$. Let us recall the parafermionic algebra of order $2j$ :
\be
f_j^{2j}f_j^++f_j^{2j-1}f_j^+f_j+\ldots+f_j^+f_j^{2j}=\alpha_jf_j^{2j-1}\label{parafermion1}
\ee
\be
f_j^{2j+1}=0\label{parafermion2}
\ee
plus the adjoint equations. For future purpose, it is interesting to introduce another algebra. We call it\cite{bes1} the pseudo-fermionic algebra of order $2j$. It is generated by $\ai^\pm$, for $1\leq i\leq 2j$, satisfying :

\be
{\ai^\pm}^2=0
\ee
\be
[\ai^\pm,\aj^\pm]=0
\ee
for $i\not=j$, and
\be
\{\aim,\aix\}=1
\ee
It is easily verified that $N_i=\aix\aim$ are number operators, so that $H_{pf}=\sum_iN_i$ is the Hamiltonian\footnote{as always, up to an additive constant}. The Green ans\"atze\cite{green} consists in taking $F_j^\pm:=\sum_i\ai^\pm$. Then the $F_j^\pm$ satisfy the defining equations of the parafermionic algebra of order $2j$, as their lowercase counterparts. This defines an algebra homomorphism of the parafermionic algebra to the pseudo-fermionic algebra. The reader must be warned that this homomorphism is not into : that is to say some equations satisfied by the $F_j^\pm$ need not be satisfied by the $f_j^\pm$. For example, for all $j$, the following equation and its adjoint, that we will call Green's equations in the sequel, are satisfied :

\be
[[F_j^+,F_j^-],F_j^-]=-2F_j\label{greeneq}
\ee

Meanwhile the $f_j^\pm$ do not satisfy Green's equations. This can be seen by first showing that the equations defining the parafermionic algebra are confluent. Then for $j>2$ equation (\ref{greeneq}) cannot hold for the $f_j^\pm$ because it only contains monomials of degree three or less. But these must be free of relations, because equations (\ref{parafermion1})-(\ref{parafermion2}) are of a greater degree. For $j=2$, it is easily seen that the left-hand side of (\ref{greeneq}) cannot be reduced to the right-hand side using (\ref{parafermion1})-(\ref{parafermion2})+adjoint. In fact, Green originally defined a parafermion to be a particle whose creation and annihilation operators satisfy Green's equations. What we have just said is that a parafermion of order $2j$ as defined by equations (\ref{parafermion1})-(\ref{parafermion2}) is not a parafermion in the sense of Green ! It is only when they are realized by the Green ans\"atze (when they are sent to the pseudofermionic algebra by some homomorphism) that the parafermionic operators of order $2j$ satisfy Green's equations, and truly become parafermionic in Green's sense (satisfy Green's equation). This is something that is not always made clear in the litterature.

\section{Number Operator Algebras}
The concept of number operator algebras is an attempt to take a new look at the quantization of the harmonic oscillator. Classically, a system of $n$ uncoupled harmonic oscillators\footnote{as the reader must now have noticed, we take all constants equal to unity} satisfy :

\be
\{H_i,p_j\}=\delta_{ij}q_j
\ee
\be
\{H_i,q_j\}=-\delta_{ij}p_j
\ee

where $H_i={1\over 2}(p_i^2+q_i^2)$ and the conjugate momenta satisfy :

\be
\{p_i,p_j\}=\{q_i,q_j\}=0\label{classcr1}
\ee
\be
\{p_i,q_j\}=\delta_{ij}\label{classcr2}
\ee
Here we use the ordinary Poisson bracket but it could be replaced with an $\epsilon$-P.B. where $\epsilon$ is a more general commutation factor, giving rise to an $\epsilon$-classical system\cite{bes2}. The usual quantization procedure goes as follows : one replaces the position and momemtum by their quantum counterparts in the classical Hamiltonian, to get :

\be
H^i_{quant}={1\over 2}(p_i^2+q_i^2)
\ee

or

\be
H^i_{quant}=\aix\aim+\mbox{constant}
\ee

when we introduce creation and annihilation operators. These will satisfy canonical commutation relations (CCR) or canonical anti-commutation relations (CAR) for fermions. If one agrees to delve into $\epsilon$-algebra, one also finds the pseudo-bosons\footnote{These are anticommuting bosons, just as the pseudo-fermions are commuting fermions} and pseudo-fermions, satisfying $\epsilon$-commtutation relations\cite{bes2}. Introducing the number operators $N_i=\aix\aim$, we can easily verify the equations 

\be
[N_i,\aj^\pm]=\pm\delta_{ij}\aj^\pm\label{noa}
\ee

which entails 
\be
[H,\ai^\pm]=\pm\ai^\pm\label{qevol}
\ee

with $H=\sum_iN_i$. The last equations are interesting because they always involve a commutator, regardless of which statistics we have adopted. Now the idea is to consider (\ref{qevol}) as the true quantum equation that we seek to obtain. We allow for any commutation relation between the creation and annihilation operators. Thus we define a number operator algebra (NOA) in the following way :

\begin{definition}
A NOA of type $n$ is an algebra generated by $a_1^-,\ldots,a_n^-,a_1^+,\ldots,a_n^+$ with an involution exchanging $\aim$ and $\aix$, and with $n$ elements $N_1,\ldots,N_n$ such that the equations (\ref{noa}) are satisfied.
\end{definition}

Note that the $N_i$ will necessarily commute with each other (see \cite{bes1}). 
We will say that a NOA is quadratic if it is defined by relations of degree at most two between the generators. It will be symmetric if a permutation of the indices of the generators induces an automorphism of the algebra. Note that the algebra of $n$ bosons, $n$ fermions, $n$ pseudo-fermions or pseudo-bosons are all symmetric and quadratic NOA. The number operators are given by $N_i=\aix\aim$ in each case, and are unique up to an additive constant. The same is true for the algebra of orthofermions of order $n$. Now the theorem is that these are essentially the only examples :

\begin{theorem}\label{th1}
Up to a rescaling of the generators and an involution, the only quadratic and symmetric NOA of type $n$, such that its defining relations are confluent in the sense of Bergman\footnote{for any ordering} are :
the algebra of bosons, fermions, pseudo-bosons, pseudo-fermions, and orthofermions, of order $n$.
\end{theorem}

When we have first proved this theorem\cite{bes3,bes4}, we were completely unaware of any previous appearance of pseudo-bosons/fermions or orthofermions in the litterature (the last ones seem to be little known anyway). Thus with our point of view on the harmonic oscillator, we have been able to independently rediscover many of the particles following exotic statistics that had been studied previously for their own sake. Thus number operator algebras serve as a unifying concept. Note that in ref. \cite{mishraraja1,mishraraja2}, 
another unifying theory is proposed, making use of generalized Fock spaces.

The reader may notice that theorem \ref{th1} does not refer to other well-known particles of exotic statistics, namely para-fermion, para-bosons and $q$-bosons. A remark is in order here.

The algebra of para-fermions/bosons is clearly a NOA thanks to Green's equations, with number operators $N_i={1\over 2}[F_i^+,F_i^-]$ for para-fermions, $N_i={1\over 2}\{F_i^+,F_i^-\}$. But this NOA is generated by cubic relations. Nevertheless, by the Green ans\"atze one asks for simpler (quadratic) relations to generate Green's equations. In this case one falls into the scope covered by our theorem : para-fermions can be realized using pseudo-fermions, and para-bosons using pseudo-bosons. Interestingly enough, one can alternatively use orthofermions\cite{mosta2} to construct parafermions.

The algebra of $q$-bosons is indeed taken care of by our theory. Recall that a $q$-boson creation and annihilation operators satisfy :

\be
a^-a^+-qa^+a^-=1
\ee
 and that the corresponding number operator (or Hamiltonian) is\cite{bes4,cs} :

\be
N=\sum_{k=0}^\infty {(1-q)^k\over 1-q^k}\aix^k\aim^k\label{qnumb}
\ee

This means that the algebra generated by $a^\pm$ should contain series such as (\ref{qnumb}). A possibility is that the algebra is topologically generated, in a certain sense. Then, for $n=1$ degree of freedom, the algebra of one $q$-boson is the only quadratic NOA apart from the usual boson and fermion case. The theory is easily extended for $n$ infinite. In this case one finds $q$-bosons and pseudo-$q$-bosons as the only new cases (note that orthofermions disappear). See ref. 
\cite{bes1,bes3,bes4} for details.

\section{A note about quasi-orthofermions}

In this section we use a new NOA to define a new type of generalized SUSY. For $\beta\not=0,1$, the algebra $Q_\beta$ generated by $a_1^\pm$, $a_2^\pm$ subject to the relations :

\be
\aim\ajm=\aix\ajx=0\label{q1}
\ee
\be
a_1^-a_1^++a_2^-a_2^++\beta a_1^+a_1^-+\beta a_2^+a_2^-=1\label{q2}
\ee

is a symmetric and quadratic NOA. Note that one can easily extends this definition for more than two generators, but then it would not contain number operators. The relations (\ref{q1})-(\ref{q2}) immediately entail :

\be
a_2^+a_1^+a_1^-=a_1^-a_1^+a_2^-+({1\over\beta}-1)a_2^-
\ee
\be
a_1^-a_2^+a_2^-=-a_1^-a_1^+a_1^-+{1\over\beta}a_1^-
\ee
\be
a_2^+a_2^-a_1^+=-a_1^+a_1^-a_1^++{1\over\beta}a_1^+
\ee
\be
a_2^+a_1^-a_1^+=a_1^+a_1^-a_2^+-({1\over\beta}-1)a_2^+
\ee

Together with the relations (\ref{q1})-(\ref{q2}) they form a confluent system for the degree-lexicographic ordering generated by $a_1^+<a_2^+<a_1^-<a_2^-$. Thus $Q_\beta$ is linearly spanned by those monomials not containing any of $\aim\ajm$, $\aix\ajx$, $a_2^-a_2^+$, $\aim\ajx\ajm$ (with $i\not=j$), $a_2^+a_2^-a_1^+$, $a_2^+a_1^-a_1^+$ as a subword. The number operators are :

\be
N_1={\beta\over\beta-1}(a_1^-a_1^++\beta a_1^+a_1^-)+\beta a_2^+a_2^-
\ee
and
\be
N_2=-{\beta\over\beta-1}(a_1^-a_1^++a_1^+a_1^-)
\ee
and the Hamiltonian is 
\be
H_{qo}=\beta(a_1^+a_1^-+a_2^+a_2^-)
\ee

Let $b^\pm$ be usual bosonic operators and let $Q_i:=b^+\aim$ ($i=1,2$) be the supercharge. It is immediate that $Q_iQ_j=Q_i^+Q_j^+=0$. We also have :

\begin{eqnarray}
Q_1Q_1^++Q_2Q_2^++\beta(Q_1^+Q_1+Q_2^+Q_2) &=& b^+b^-(a_1^-a_1^++a_2^-a_2^+)+\beta b^-b^+(a_1^+a_1^-+a_2^+a_2^-)\nonumber \\
 &= & b^+b^-(1-\beta a_2^+a_2^--\beta a_1^+a_1^-)+\ldots\nonumber\\
&=& b^+b^-+\beta(b^-b^+-b^+b^-)(a_1^+a_1^-+a_2^+a_2^-)\nonumber\\
&=& H_b+H_{qo}\nonumber\\
&:=&H\nonumber\\
\end{eqnarray}

Then we have automatically $[H,Q_i]=[H,Q_i^+]=0$. Thus one can define the ``quasi-orthoSUSY'' algebra of order $n$ by :

\be
Q_iQ_j=Q_i^+Q_j^+=0
\ee
\be
\sum_{i=1}^n Q_iQ_i^++\beta\sum_{i=1}^nQ_i^+Q_i=H
\ee
\be
[H,Q_i]=[H,Q_i^+]=0
\ee

We have just realized a quasi-orthosupersymmetric system of order two for $\beta\not=0,1$. For $n=1$, $\beta=1$ we recover the usual $N=1$ SUSY algebra.
\section{Concluding remarks}

In the quantum theory, the concept of a particle is intimately tied with that of a harmonic oscillator. More precisely, the kind of statistics a particle belongs to is determined by the commutation relations of its creation and annihilation operators, what in turn can be viewed as a quantization of the Poisson-Bracket relation between the canonical momenta of a harmonic oscillator. Since we have undertaken to find all these quantizations (subject to some combinatorial restrictions) that entails the quantum equation of evolution (\ref{qevol}), it should not be surprising that we have rediscovered various particles of exotic statistics, as well as a new one (quasi-orthofermion). However, we hope to have shown the interest of our systematic combinatorial approach.

\end{document}